\documentclass[
aps,prd,
12pt,
nopreprintnumbers,
showpacs,
eqsecnum,
nofootinbib
]{revtex4-1}

\usepackage{graphicx}

\def\v0{{\bf 0}}

\begin{document}

\title{
Fermion spectrum and localization on kinks in a deconstructed dimension
}
\author{Nahomi Kan}\email[]{kan@yamaguchi-jc.ac.jp}
\affiliation{
Yamaguchi Junior College,
Hofu-shi, Yamaguchi 747--1232, Japan}
\author{Koichiro Kobayashi}\email[]{m004wa@yamaguchi-u.ac.jp}
\author{Kiyoshi Shiraishi}\email[]{shiraish@yamaguchi-u.ac.jp}
\affiliation{
Yamaguchi University,
Yamaguchi-shi, Yamaguchi 753--8512, Japan}
\date{\today}

\begin{abstract}
We study the deconstructed scalar theory having nonlinear interactions
and being renormalizable. It is shown that the kink-like configurations
exist in such models. The
possible forms of Yukawa coupling are considered. We find the degeneracy
in mass spectrum of fermions coupled to the nontrivial scalar
configuration.
\end{abstract}


\pacs{
05.45.Yv, 
11.10.Kk, 
11.10.Lm, 
11.27.+d  
}

\maketitle

\section{Introduction}

The concept of the brane and brane world exhibits broad possibilities in
the particle theory beyond the standard model.
In the early stage of study on the higher dimensional theories,
it has been considered that we may live in a domain wall perpendicular
to the fifth dimension~\cite{dw}.
The domain wall may be described by a kink configuration of an
self-interacting scalar field. The matter field which couples to the kink
is localized in the extra dimension and acquires a mass spectrum
consisting both discrete and continuum modes in general.

Even if we consider the model with a compact dimension,
the kink or solitonic configuration is expected to play important roles.
The localization of fermions due to solitonic domain walls can explain the
hierarchy in masses and the Yukawa couplings~\cite{ex}. The compactified
dimension is taken as $S^1$ or $S^1/Z_2$ in such a case.
The kink solutions in the periodic compact space are known to be
topologically  unstable like sphalerons. They have been studied by some
authors~\cite{sph}.

In the recent decade, another approach to the particle physics models has
been explored. The idea of dimensional deconstruction~\cite{DD} is to
construct the models mimicking the higher dimensional gauge theory from
the four dimensional gauge theory. In some sense, the deconstruction means
introducing the discrete extra space into the theory. Some merits of
higher dimensional theory can be inherited by four dimensional theory,
whose high-energy behavior can be well controlled.

Therefore it is worth examining the `deconstructed kink
model' in the framework of four dimensional theory.
Fortunately, discrete models for kink-like solutions  have been
considered apart from particle physics~\cite{Speight}.
Thus we must study the solutions with specific boundary conditions
and especially coupling to fermions to apply the discrete models to
particle physics.

Most general starting point is to consider the multi-scalar model in four
dimensions, whose Lagrangian density reads
\begin{equation}
{\cal L}=\frac{1}{2}\sum_{i=1}^N
\partial^\mu\phi_i\partial_\nu\phi_i-V(\phi_1,\phi_2,\ldots,\phi_N)\,,
\end{equation}
where $\phi_i$ are real scalar fields and $V$ is the potential.
Unlike the original dimensional deconstruction scheme, the gauge
invariance is not assumed. 
Moreover even if the continuum limit is uniquely defined, it is known
that the discretized action can be exhibited by various different
forms~\cite{Speight}. Since the other guiding principles are needed for
detailed study, we will refer the supersymmetric model as shown later.

The original dimensional deconstruction is found to be associated
with the graph structure~\cite{KS}. Thus we begin with considering the
bilinear term in scalar fields and construct the interaction terms on
vertices or edges of a graph taking the correspondence with the continuum
theory into consideration.

In the next section, we review the kink solution in $(1+1)$-dimensional
field theory. Since we must include fermions later, we consider a
supersymmetric theory; the structure of this theory will be referred in
the later sections. In Sec.~\ref{sec3}, we consider discretized or
deconstructed free scalar models, whose continuum limit is the
compactified theory with
$S^1$ or
$S^1/Z_2$. In Sec.~\ref{sec4}, we choose the form of the scalar
interactions, which is expected to make the `discrete kink'
configurations possible. The result of numerical calculations on the
solitonic solutions is shown in Sec.~\ref{sec5}.  In Sec.~\ref{fermion},
we consider the Yukawa coupling of fermions and scalars. Various
`discrete' versions of the Yukawa coupling are possible, though their
`continuum' limits are all the same one. In this paper, we consider the
models whose couplings are similar to those of the supersymmetric model
mentioned in Sec.~\ref{kink} in the continuum limit. The numerical result
for the fermionic spectra is exhibited in Sec.~\ref{fmass}. The last
section is devoted to summary and prospects.


\section{kink solutions in the continuum theory
\label{kink}}
In this section, we review a continuum theory for kink solutions
\cite{Raj}. The form of the action is used to determine the way for
discretization of bosonic as well as fermionic sector later.

We consider the
$(1+1)$-dimensional supersymmetric field theory, whose Lagrangian density
is
\begin{equation}
{\cal L}=\frac{1}{2}(\partial\phi)^2-\frac{1}{2}
\left(\frac{\partial W(\phi)}{\partial\phi}\right)^2
+\frac{1}{2}i\bar{\psi}\partial\!\!\!/
\psi-\frac{1}{2}\frac{\partial^2 W(\phi)}{\partial\phi^2}\bar{\psi}\psi\,,
\end{equation}
where we choose the superpotential $W(\phi)$ as
\begin{equation}
W(\phi)=g\left(\frac{\phi^3}{3}-a^2\phi\right)\,,
\end{equation}
with $g$ and $a$ are constants.

Then the bosonic part of the Lagrangian density reads
\begin{equation}
{\cal L}_B=\frac{1}{2}(\partial\phi)^2-\frac{1}{2}g^2
\left(\phi^2-a^2\right)^2\,,
\end{equation}
and thus we find the equation of motion as
\begin{equation}
-\partial_t^2\phi+\partial_x^2\phi-2g^2(\phi^2-a^2)\phi=0\,.
\end{equation}

A non trivial static solution, a kink solution, satisfies the equation of
motion, and expressed as
\begin{equation}
\phi_0(x)=a\tanh g a x\,,
\end{equation}
where we set an integration constant for $\phi_0(0)=0$.
One can find that this solution also satisfies
\begin{equation}
\partial_x\phi+g
\left(\phi^2-a^2\right)=\partial_x\phi+\frac{\partial W}{\partial
\phi}=0\,.
\label{BPS}
\end{equation}



The coupling of the scalar and the (Majorana) fermion is proportional  to
$-g\phi\bar{\psi}\psi$. In the present case, it is not surprising
 and is of the usual Yukawa form. This coupling with general coupling
constant (say, $-G\phi\bar{\psi}\psi$) or similar type of models have been
investigated in the soliton physics~\cite{DJR}.


The solitonic configurations in the periodic compact space have also 
been considered in several authors~\cite{sph}.
It has been shown that he nature of quasi-stable solutions depends on the
size of the compact space.
A solution of (\ref{BPS}) is not necessarily a solution of the equation of
motion in the compact space. Interestingly, these features are also true
in some sense for the case of discrete solitons studied in
Sec.~\ref{sec4}.

\section{deconstructing a free scalar field theory and a graph
\label{sec3}}
Suppose $N$ real scalar fields $\phi_i$ $(i=1,\dots, N)$.
The mass spectrum of the scalars is the same as that of the vector
bosons in the dimensional deconstruction model~\cite{DD},
if the action is given by
\begin{eqnarray}
{\cal
L}&=&\sum_{i=1}^N\frac{1}{2}(\partial\phi_i)^2-\frac{1}{2}f^2
\sum_{i}^{}({{\it\Delta}}\phi_{i})^2\nonumber \\
&=&\sum_{i=1}^N\frac{1}{2}(\partial\phi_i)^2-\frac{1}{2}f^2
\sum_{i}^{}(\phi_{i+1}-\phi_i)^2\,,
\label{mm}
\end{eqnarray}
where ${\it\Delta}\phi_i\equiv\phi_{i+1}-\phi_i$ and $f$ is a mass scale.
In the four-dimensional point of view, we can say that the potential is of
a `difference-squared' type.

The sum in the potential is interpreted as $\phi_{N+1}\equiv\phi_1$ and
from $i=1$ to $N$. Then the large $N$ limit leaving $L\equiv N/f$ constant
leads to the five dimensional scalar theory with $S^1$ (Kaluza-Klein)
compactification.
Namely, the $N\times N$ mass-square matrix in this model is
\begin{equation}
f^2\left(
\begin{array}{cccccc}
2 & -1 & 0 & \cdots & 0 & -1\\
-1 & 2 & -1 & \cdots & 0 & 0\\
0 & -1 & 2 & \cdots & 0 & 0\\
\vdots & \vdots & \vdots &\vdots & \vdots & \vdots\\
0 & 0 & 0 & \cdots & 2 & -1 \\
-1 & 0 & 0 &\cdots & -1 &2
\end{array}
\right)\equiv f^2{\Delta}(C)\,,
\end{equation} 
and the eigenvalues are 
\begin{equation}
4f^2\sin^2\frac{k\pi}{N}\qquad (k=0, 1, \ldots, N-1)\,,
\end{equation}
which becomes $(2\pi k/L)^2$ in the large $N$ limit mentioned above.
Here $L$ corresponds to the circumference of $S^1$.

The other summation rule is possible. The sum in the
expression~(\ref{mm}) is taken from $i=1$ to $N-1$. In this case 
the $N\times N$ mass-square matrix becomes
\begin{equation}
f^2\left(
\begin{array}{cccccc}
1 & -1 & 0 & \cdots & 0 &  0\\
-1 & 2 & -1 & \cdots & 0 & 0\\
0 & -1 & 2 & \cdots & 0 & 0\\
\vdots & \vdots & \vdots &\vdots & \vdots & \vdots\\
0 & 0 & 0 & \cdots & 2 & -1 \\
0 & 0 & 0 &\cdots & -1 & 1
\end{array}
\right)\equiv f^2{\Delta}(P)\,,
\end{equation} 
and its eigenvalues are 
\begin{equation}
4f^2\sin^2\frac{k\pi}{2N}\qquad (k=0, 1, \ldots, N-1)\,,
\end{equation}
and the large $N$ limit yields the theory with the compactification on
$S^1/Z_2$.

In conclusion so far, 
the construction of the potential is associated with the structure of a
graph, in the sense of graph theory~\cite{graphtheory}.
A graph consists of vertices and edges each of which connects two
vertices. The edge $e$ connects the vertices $o(e)$ and $t(e)$. 
The vertex $o(e)$ means the origin of the edge $e$
while the vertex
$t(e)$  means the terminus of the edge $e$. 
We assign
$N$ scalar fields onto
$N$ vertices of the graph and their action is written by 
\begin{equation}
{\cal
L}=\sum_{v\in {\cal V}}\frac{1}{2}(\partial\phi_v)^2-\frac{1}{2}f^2
\sum_{e\in {\cal E}}(\phi_{t(e)}-\phi_{o(e)})^2\,,
\end{equation}
where ${\cal V}$ and ${\cal E}$ denote the set of vertices and edges,
respectively.

Now we can express the mass term by using the incidence matrix
$E$ defined as
\begin{equation}
(E)_{ve}=\left\{
\begin{array}{cc}
1 & {\rm if~}v=o(e)\\
-1 & {\rm if~}v=t(e)\\
0 & {\rm otherwise}
\end{array}
\right.\,,
\label{Ei}
\end{equation}
since
in both cases, the difference ${\it\Delta}\phi$ is defined as associated
with each edge such that
\begin{equation}
({\it\Delta}\phi)_e=\phi_{t(e)}-\phi_{o(e)}=-\sum_{v\in
{\cal V}}(E^T)_{ev}\phi_v\,,
\end{equation}
and then
\begin{equation}
\sum_{e\in {\cal E}}(\phi_{t(e)}-\phi_{o(e)})^2=
\sum_{v,v'\in {\cal V}}\sum_{e\in
{\cal E}}\phi_{v'}(E)_{v'e}(E^T)_{ev}\phi_v\,,
\end{equation}
where the matrix $E^T$ is the transposed matrix of $E$.

\begin{figure}[ht]
\centering
\includegraphics[height=4cm]
{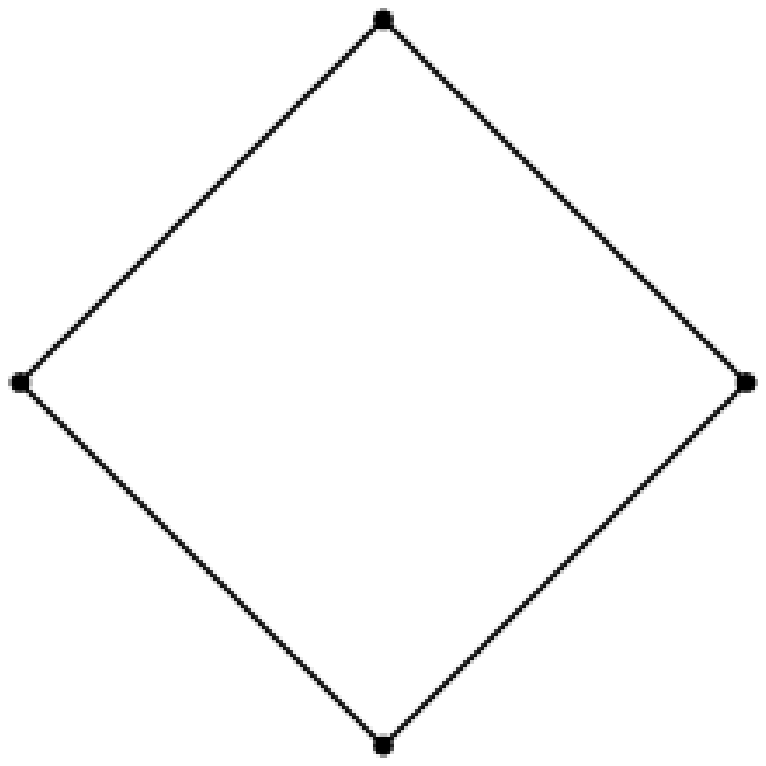}
\qquad\qquad
\includegraphics[height=4cm]
{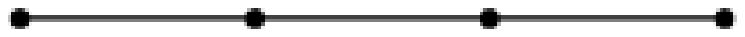}
\caption{%
The cycle graph $C_4$ and the path graph $P_4$.
Each suffix indicates the number of vertices in the graph.}
\label{CP}
\end{figure}

For the first case introduced in this section, the incidence matrix
is
\begin{equation}
E(C)=\left(
\begin{array}{ccccc}
1& 0& \cdots&0& -1\\
-1&1& \cdots&0& 0\\
0& -1& \cdots& 0& 0\\
\vdots &\vdots &\ddots &\vdots&\vdots \\
 0& 0& \cdots&1& 0\\
 0& 0& \cdots&-1&1
\end{array}
\right)\,,
\end{equation}
and its transposed matrix is
\begin{equation}
E^T(C)=\left(
\begin{array}{cccccc}
1& -1& 0& \cdots& 0 &0\\
0& 1& -1& \cdots& 0 &0\\
\vdots& \vdots&\vdots &\ddots &\vdots &\vdots\\
0& 0& 0&\cdots &1&-1\\
-1& 0& 0&\cdots &0&1
\end{array}
\right)\,.
\end{equation}
The corresponding graph is the {\it cycle graph} in the term of graph
theory (Fig.~\ref{CP}).
Note that $\Delta(C)=E(C)E^T(C)$,
where $\Delta(C)$ is known as a graph Laplacian for a cycle graph.

For the second case introduced in this section, the incidence matrix
is
\begin{equation}
E(P)=\left(
\begin{array}{cccc}
1& 0& \cdots&0 \\
-1& 1& \cdots&0 \\
0& -1& \cdots& 0\\
\vdots &\vdots &\ddots &\vdots \\
 0& 0& \cdots&1 \\
 0& 0& \cdots&-1
\end{array}
\right)\,,
\end{equation}
and its transposed matrix is
\begin{equation}
E^T(P)=\left(
\begin{array}{cccccc}
1& -1& 0& \cdots& 0 &0\\
0& 1& -1& \cdots& 0 &0\\
\vdots& \vdots&\vdots &\ddots &\vdots &\vdots\\
0& 0& 0&\cdots &1&-1
\end{array}
\right)\,.
\end{equation}
The corresponding graph is the {\it path graph} in the term of graph
theory (Fig.~\ref{CP}).
Note that $\Delta(P)=E(P)E^T(P)$,
where $\Delta(P)$ is known as a graph Laplacian for a path graph.

In the next section, we will examine the nonlinear interaction of scalars,
which is essential for the kink-like configurations, 
by referring the graph structure.

\section{deconstructing the scalar potential
\label{sec4}}

In the previous section, 
the term corresponding to the kinetic term of the scalar
theory has been simply discretized. However, we should consider some
constraints to the interaction term in the discretized action, because
there are many possible descretizations which have the same continuum
limit.

First of all, we want to avoid apparently complicated interactions.
We postulate renormalizability, at least superficially. At most four
scalar fields coupled mutually at a point in four dimensions.
Secondly, we wish to consider models possessing a continuum limit for a
large number of four-dimensional scalar field. In this limit, the static
configuration is much alike kinks or sphalerons on $S^1/Z_2$ or $S^1$.
Thirdly, the structure of the interaction should reflect a graph
structure. The deconstructed `kinetic term' is expressed as a sum over
edges of a graph as shown in the previous section, so the deconstructed
interaction term should also be written as a sum of the contribution
assigned at edges.

Now the possible Lagrangian density is the following:
\begin{equation}
{\cal
L}=\sum_{v\in
{\cal V}}\frac{1}{2}(\partial\phi_v)^2-\frac{1}{2}f^2\sum_{e\in
{\cal E}}(\phi_{t(e)}-\phi_{o(e)})^2-
\frac{1}{2}\sum_{e\in {\cal E}}g^2\left(
\frac{p\phi_{t(e)}^2+q\phi_{t(e)}\phi_{o(e)}+p\phi_{o(e)}^2}{2p+q}-a^2\right)^2\,,
\end{equation}
where $p$ and $q$ are dimensionless constants and we consider only a path
graph or a cycle graph, corresponding to the continuum theory on
$S^1/Z_2$ or
$S^1$, respectively.

For a cycle or path graph, the potential term in the sense of the
four-dimensional theory is
\begin{equation}
V=\frac{1}{2}f^2\sum_{i}(\phi_{i+1}-\phi_i)^2+
\frac{1}{2}\sum_{i}g^2\left(
\frac{p\phi_{i+1}^2+q\phi_{i+1}\phi_i+p\phi_i^2}{2p+q}-a^2\right)^2\,.
\end{equation}
Setting the derivative with respect to $\phi_k$ ($k$ is not $1$ or $N$
for the case with a path graph) to zero yields the recurrence relation
among three terms. Unfortunately, it seems that this recurrence relation
cannot be satisfied by multiple use of some recurrence relations in two
terms. Thus for a finite number of fields, similar relations to the
continuum theory like (\ref{BPS}) cannot hold.

Because this is still far from a specific model due to arbitrary
numbers $p$ and
$q$, we select the case by special conditions $p=q=1$.
This selection was adopted first by Speight and Ward (in \cite{Speight}).
In this case, we find
\begin{eqnarray}
g\left(
\frac{\phi_{i+1}^2+\phi_{i+1}\phi_i+\phi_i^2}{3}-a^2\right)
{\it\Delta}\phi_i&=&g\left(
\frac{\phi_{i+1}^2+\phi_{i+1}\phi_i+\phi_i^2}{3}-a^2\right)
(\phi_{i+1}-\phi_i)\nonumber \\
=g\left(
\frac{\phi_{i+1}^3-\phi_i^3}{3}-a^2(\phi_{i+1}-\phi_i)\right)&=&W(\phi_{i+1})-W(\phi_i)\,.
\end{eqnarray}
This relation shows that the derivative of the
superpotential $\frac{\partial W}{\partial\phi}$ is replaced by the
difference
$\frac{{\it\Delta} W}{{\it\Delta}\phi}$, or
$${\it\Delta}W=-E^TW\,,\quad {\rm where}~~W=
\left(
\begin{array}{c}
W(\phi_1)\\
W(\phi_2)\\
\vdots \\
W(\phi_N)
\end{array}\right). $$
Thus in this case, we can say that the interaction terms is related to the
graph structure.

For $p=q=1$, which we concentrate on this case hereafter,
the potential minimum is given by the simultaneous equations
\begin{eqnarray}
\frac{\partial V}{\partial
\phi_k}&=&-f^2(\phi_{k+1}-2\phi_k+\phi_{k-1})+
g^2\left(
\frac{\phi_{k+1}^2+\phi_{k+1}\phi_k+\phi_k^2}{3}-a^2\right)
\frac{\phi_{k+1}+2\phi_k}{3} \nonumber \\
& &+g^2\left(
\frac{\phi_{k}^2+\phi_{k}\phi_{k-1}+\phi_{k-1}^2}{3}-a^2\right)
\frac{2\phi_{k}+\phi_{k-1}}{3}=0\,,
\label{zenka}
\end{eqnarray}
where $k$ is not $1$ or $N$ for
the case with a path graph.
To obtain static scalar configurations,
we should solve the recursion relation  (\ref{zenka}) for a cycle graph,
and for a path graph with additional `boundary' equations,
\begin{eqnarray}
\frac{\partial V}{\partial
\phi_1}&=&-f^2(\phi_{2}-\phi_1)+
g^2\left(
\frac{\phi_{2}^2+\phi_{2}\phi_1+\phi_1^2}{3}-a^2\right)
\frac{\phi_{2}+2\phi_1}{3}=0\,,\\
\frac{\partial V}{\partial
\phi_N}&=&-f^2(-\phi_N+\phi_{N-1})+g^2\left(
\frac{\phi_{N}^2+\phi_{N}\phi_{N-1}+\phi_{N-1}^2}{3}-a^2\right)
\frac{2\phi_{N}+\phi_{N-1}}{3}=0\,.
\end{eqnarray}

\section{numerical solutions
\label{sec5}}

We can look for symmetric configurations by numerical calculations.
The recursion relation (\ref{zenka}) can be rewritten as
\begin{eqnarray}
& &-\frac{y_{k+1}-2y_k+y_{k-1}}{h^2}+
\left(
\frac{y_{k+1}^2+y_{k+1}y_k+y_k^2}{3}-1\right)
\frac{y_{k+1}+2y_k}{3} \nonumber \\
& &+\left(
\frac{y_{k}^2+y_{k}y_{k-1}+y_{k-1}^2}{3}-1\right)
\frac{2y_{k}+y_{k-1}}{3}=0\,,
\end{eqnarray}
where $y_i\equiv \phi_i/a$ and $h\equiv g a/f$.

Although the third order equations can be analytically solved,  
here we use 
{\it Mathematica} and the 
command {\tt FindMimimum} to solve the solution for potential minima,
without detailed analysis on numerical errors.
This is partly because our model is still a toy model and because
we consider only modest number of fields as a considerable field theory.

We try to find non trivial solution for
$N=4, 8, {\rm and}~ 16$ for both cases with path and cycle graphs.

\subsection{a soliton on a path graph}

\begin{figure}[ht]
\centering
\includegraphics[height=2cm]
{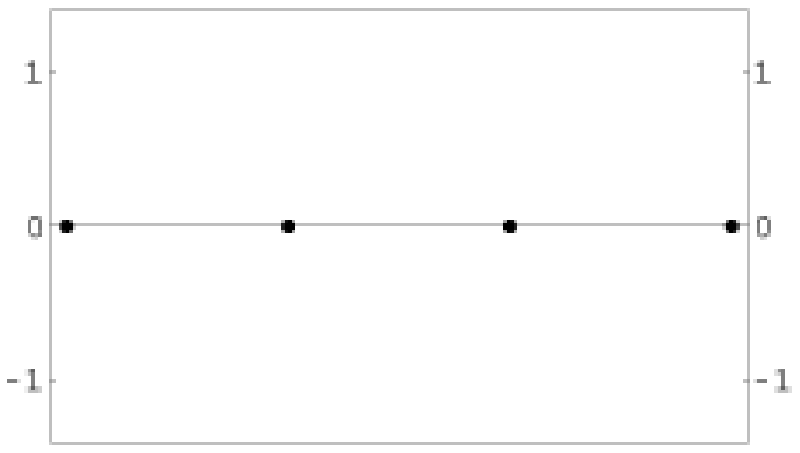}\qquad
\includegraphics[height=2cm]
{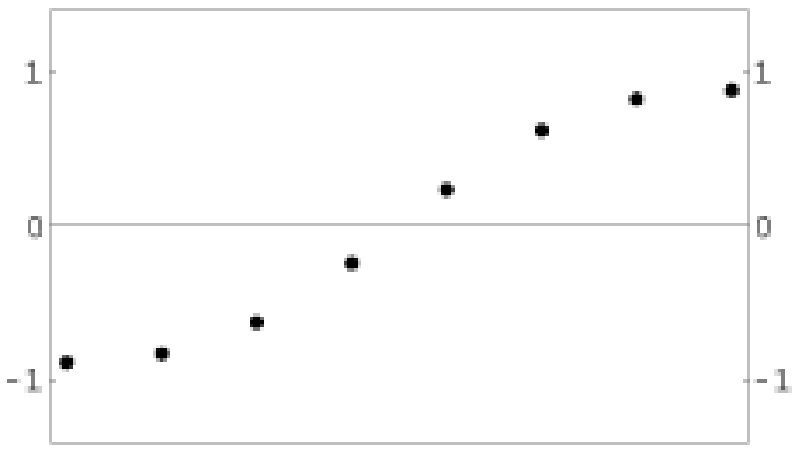}\qquad
\includegraphics[height=2cm]
{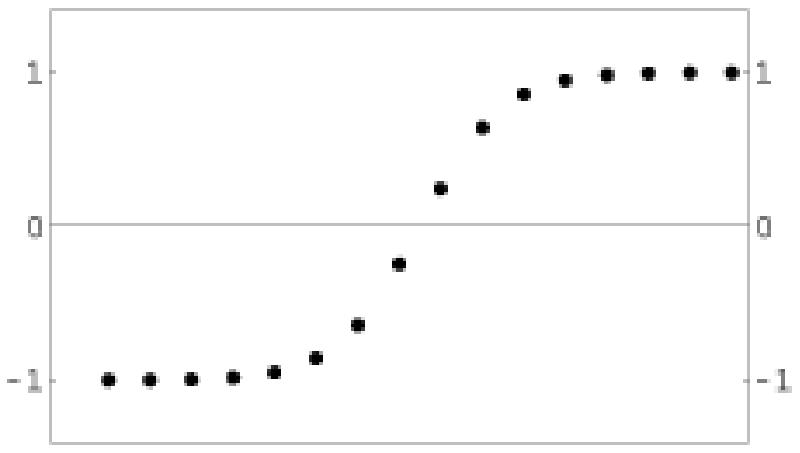}\\
\includegraphics[height=2cm]
{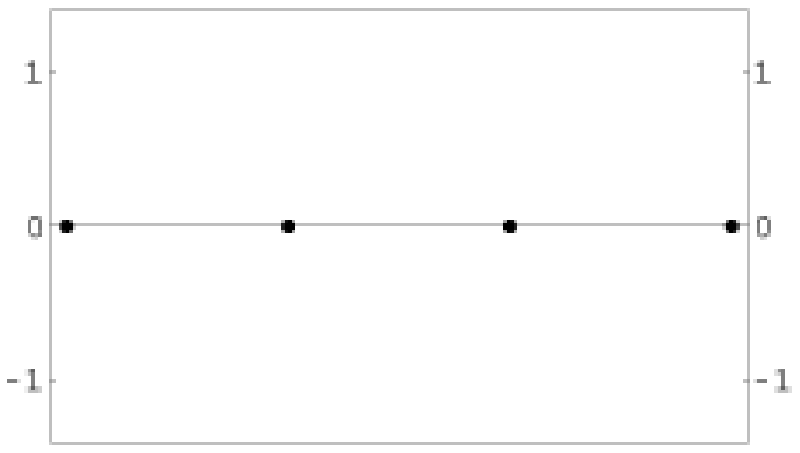}\qquad
\includegraphics[height=2cm]
{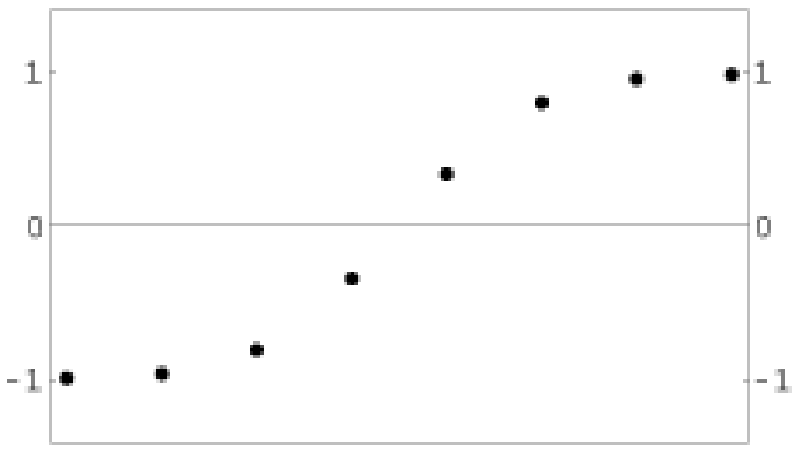}\qquad
\includegraphics[height=2cm]
{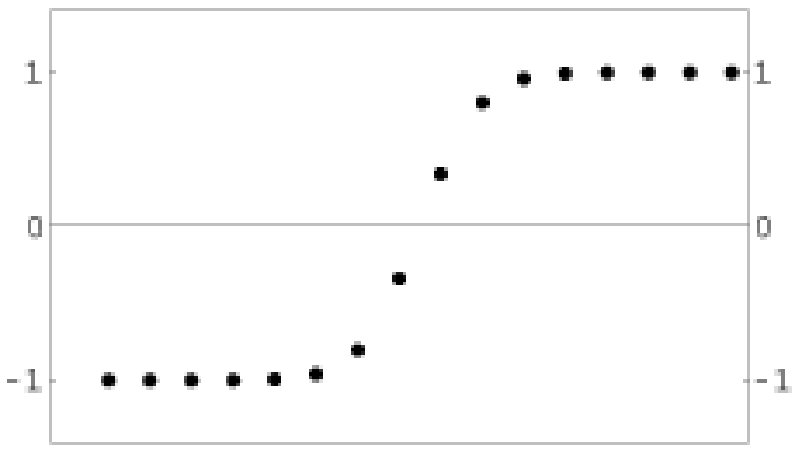}\\
\includegraphics[height=2cm]
{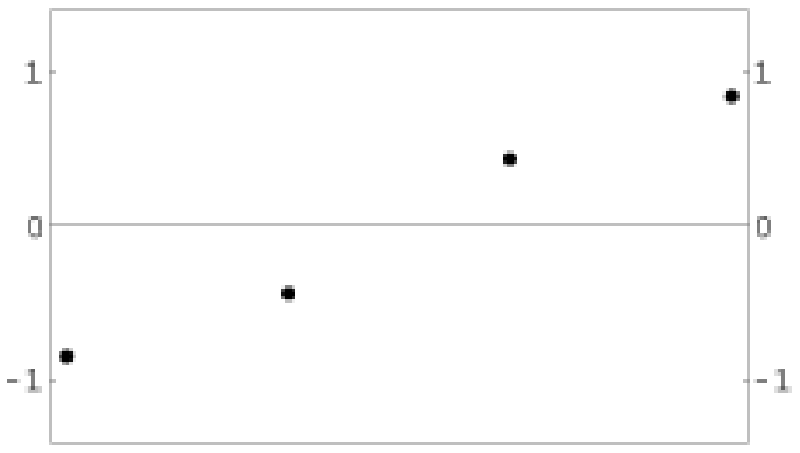}\qquad
\includegraphics[height=2cm]
{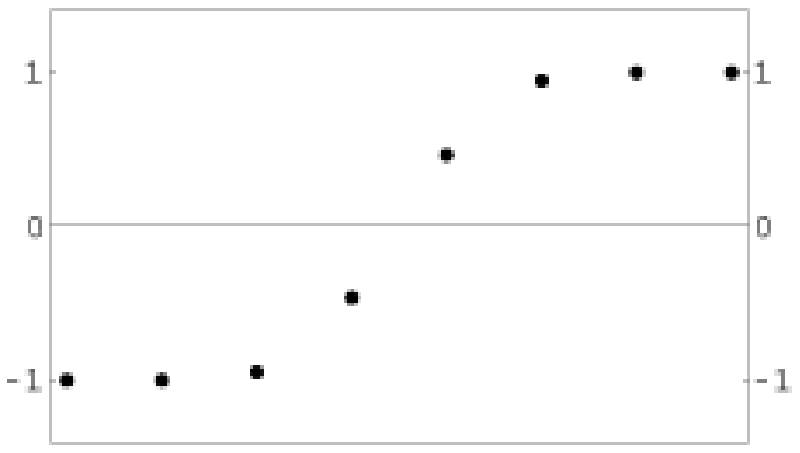}\qquad
\includegraphics[height=2cm]
{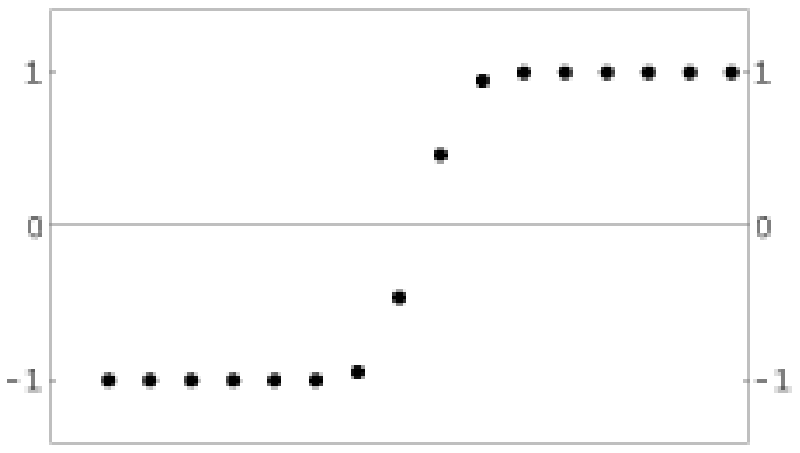}\\
\includegraphics[height=2cm]
{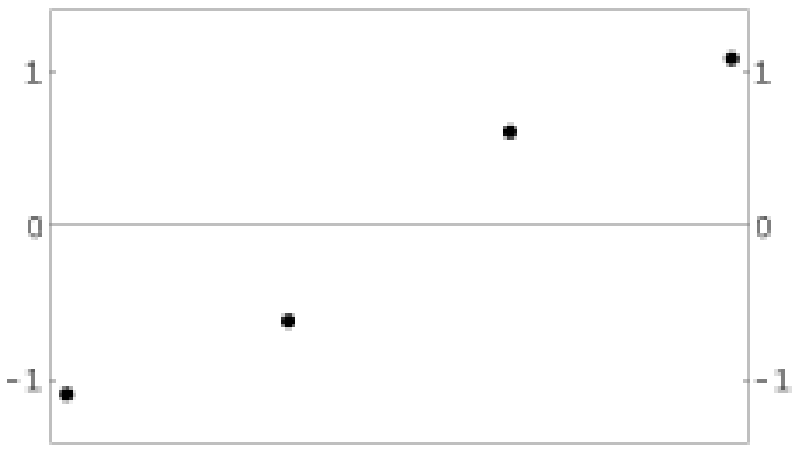}\qquad
\includegraphics[height=2cm]
{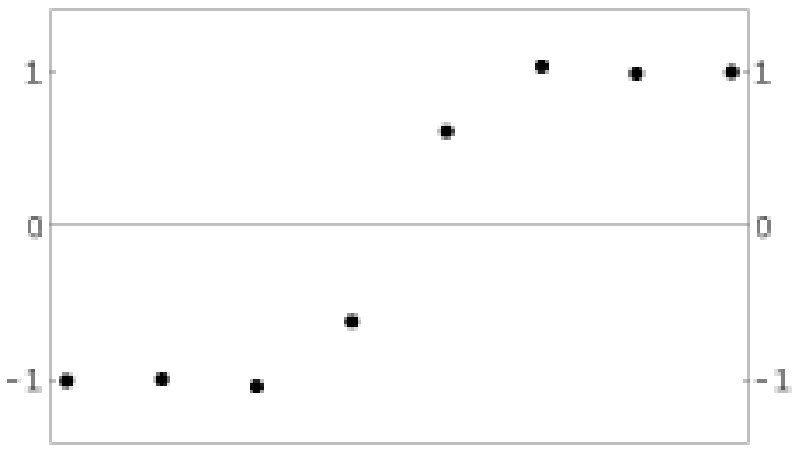}\qquad
\includegraphics[height=2cm]
{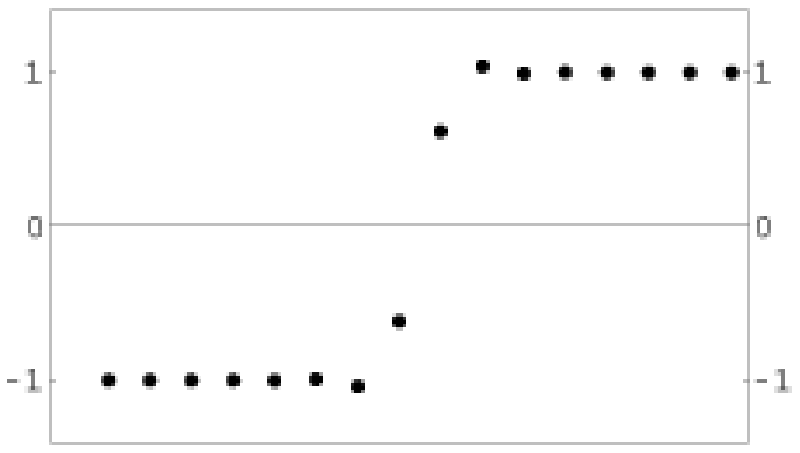}\\
\includegraphics[height=2cm]
{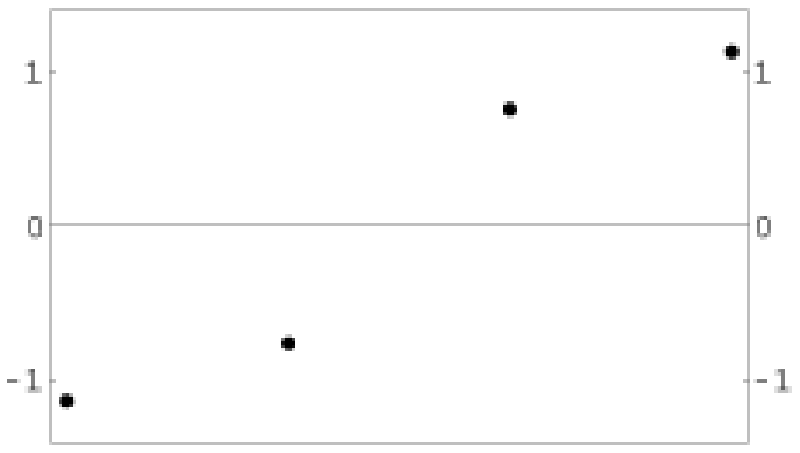}\qquad
\includegraphics[height=2cm]
{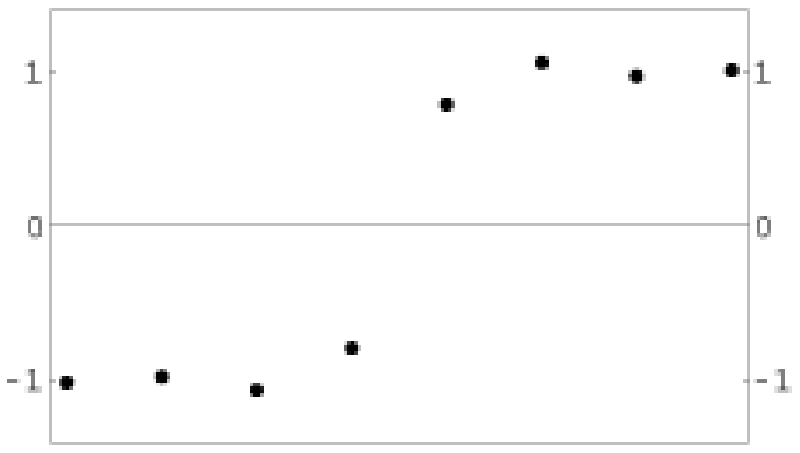}\qquad
\includegraphics[height=2cm]
{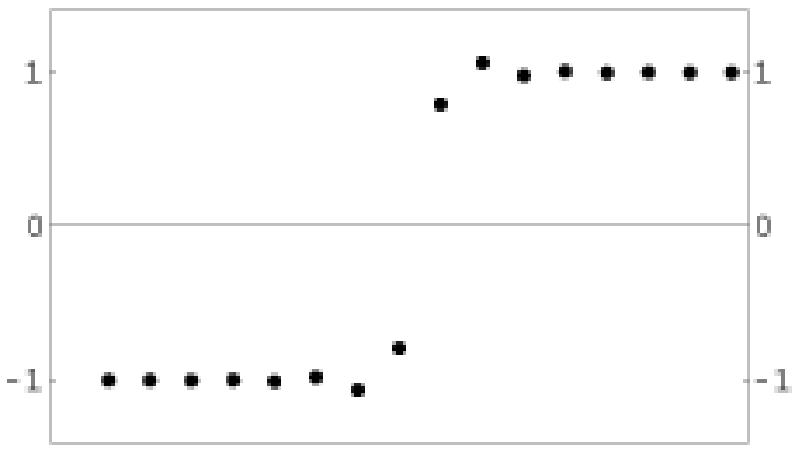}\qquad
\caption{%
Numerical solutions for the scalar field configuration 
in the cases with $P_4$ for the left column, 
$P_8$ for the center column, and
$P_{16}$ for the right column. Each vertical axis
indicates
$y_i=\phi_i/a$. The figures in the five rows correspond to the parameter
$h=2^{-1}, 2^{-0.5}, 1, 2^{0.5}, 2^1$, from the top to the bottom.}
\label{fig2}
\end{figure}

FIG.~\ref{fig2} shows the solitonic profile
for $N=4, 8,~ {\rm and}~ 16$.
Even for the small number of $N$, a non trivial configuration exists for
a sufficiently large value for $h$.
Larger values of $h$ seem to induce the excess of the value of $\phi_i$
beyond $a$, near the center of the  kink.

\subsection{a soliton on a cycle graph}

\begin{figure}[ht]
\centering
\includegraphics[height=2cm]
{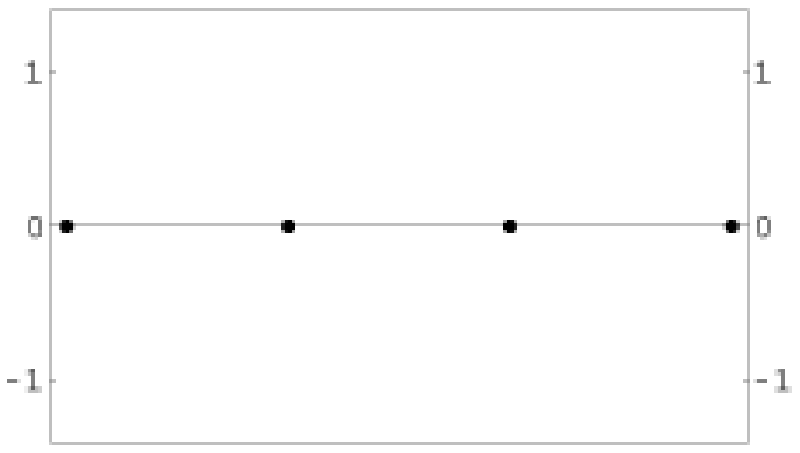}\qquad
\includegraphics[height=2cm]
{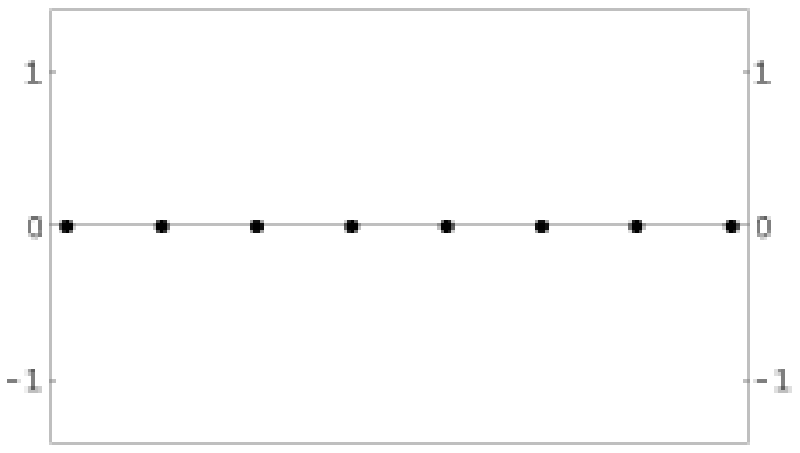}\qquad
\includegraphics[height=2cm]
{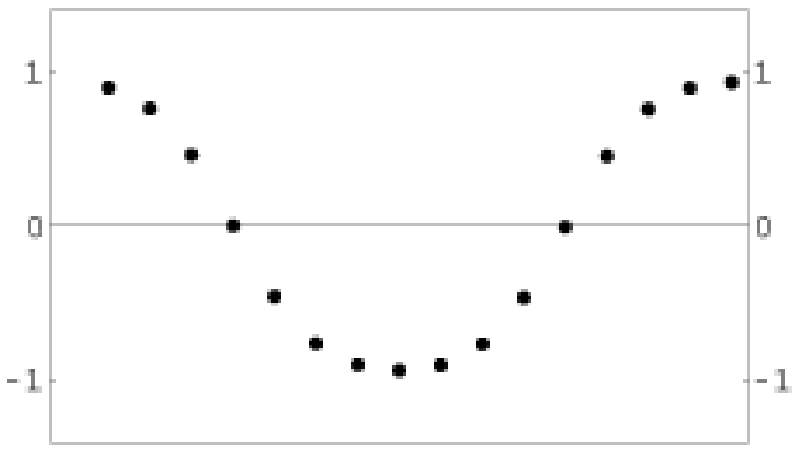}\\
\includegraphics[height=2cm]
{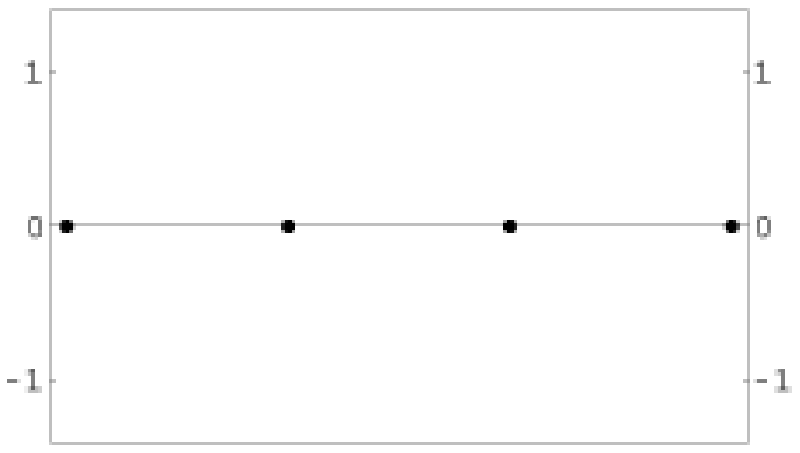}\qquad
\includegraphics[height=2cm]
{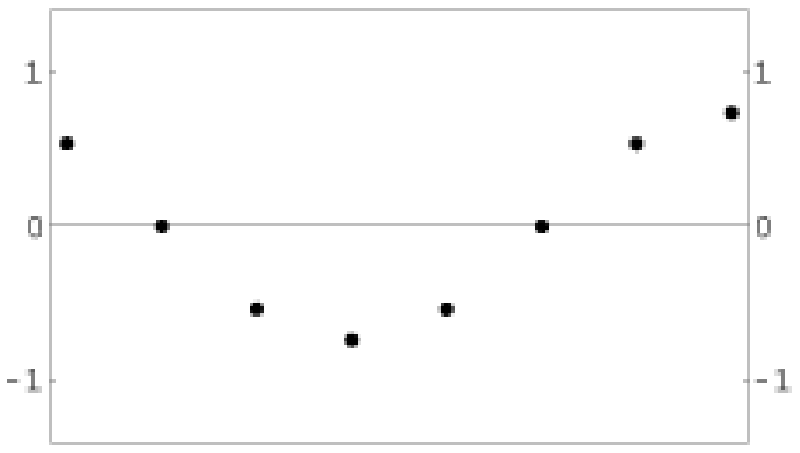}\qquad
\includegraphics[height=2cm]
{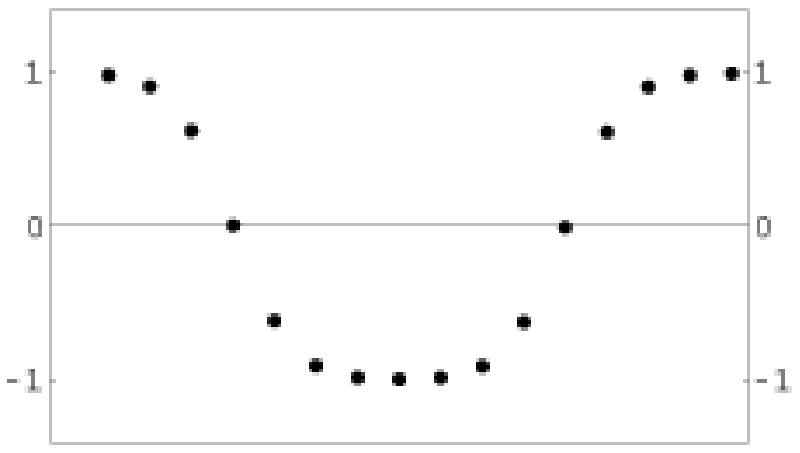}\\
\includegraphics[height=2cm]
{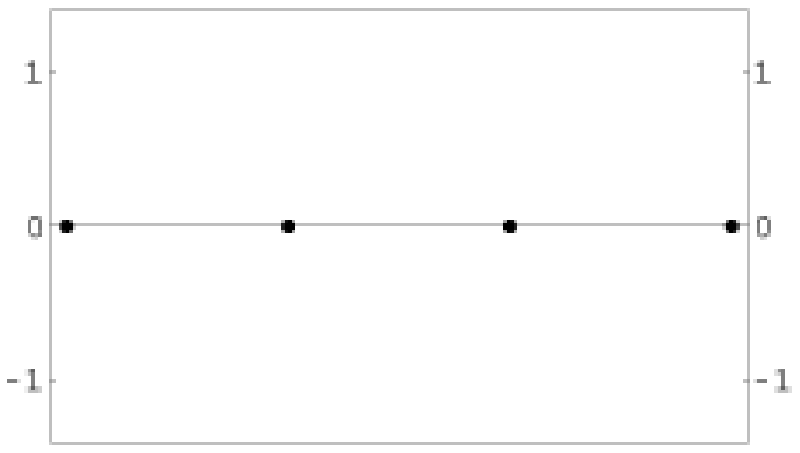}\qquad
\includegraphics[height=2cm]
{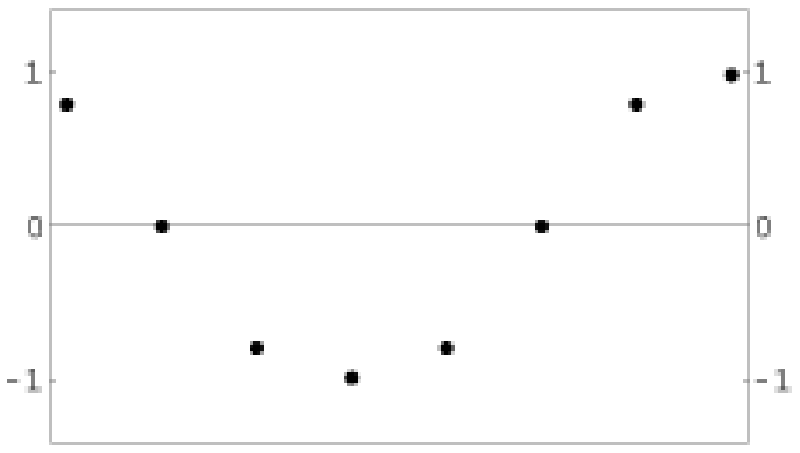}\qquad
\includegraphics[height=2cm]
{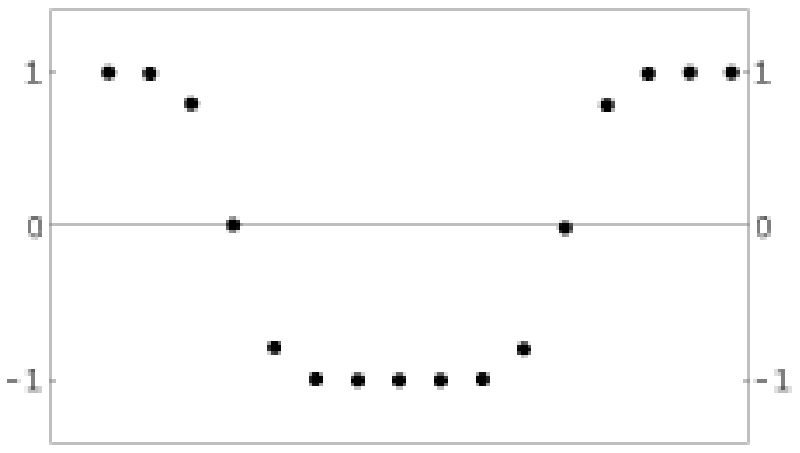}\\
\includegraphics[height=2cm]
{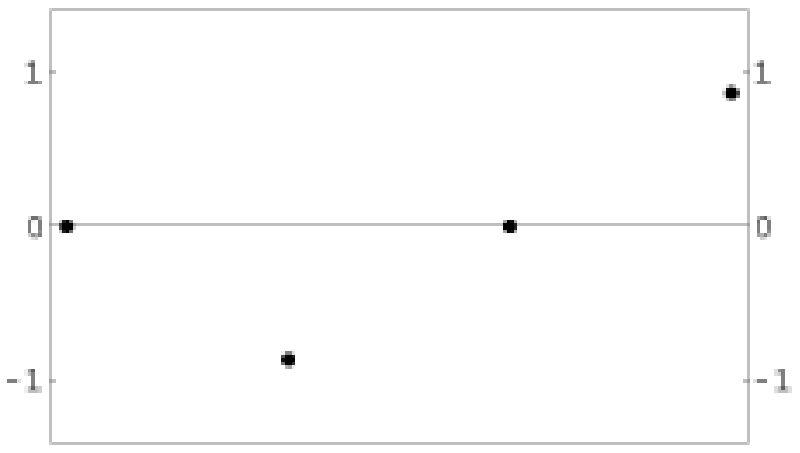}\qquad
\includegraphics[height=2cm]
{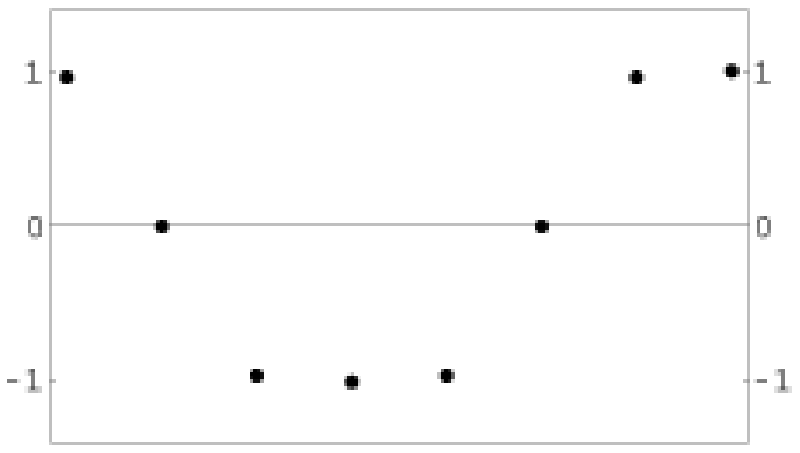}\qquad
\includegraphics[height=2cm]
{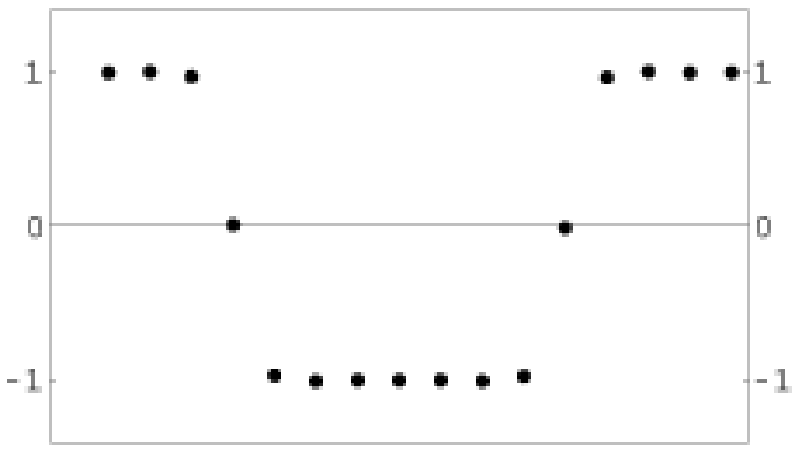}\\
\includegraphics[height=2cm]
{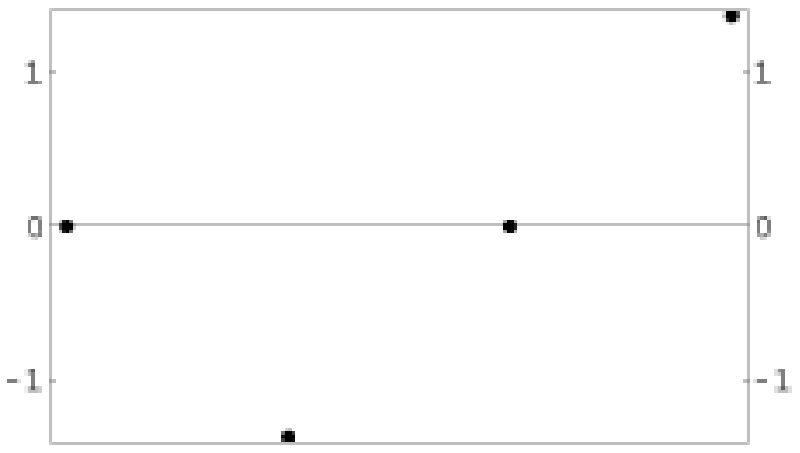}\qquad
\includegraphics[height=2cm]
{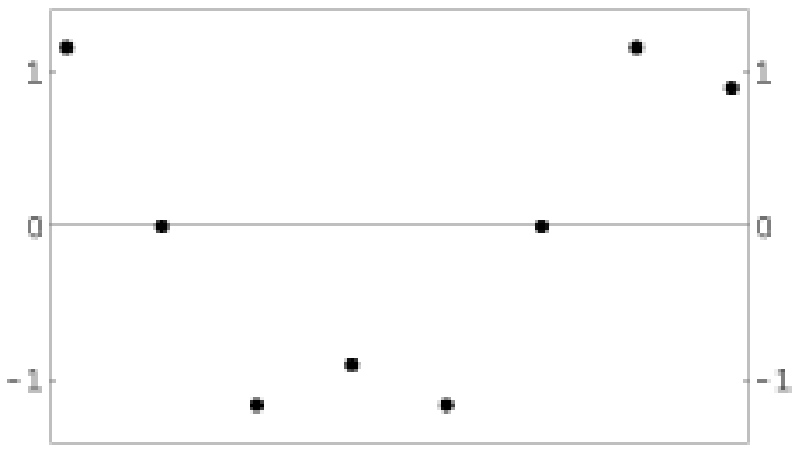}\qquad
\includegraphics[height=2cm]
{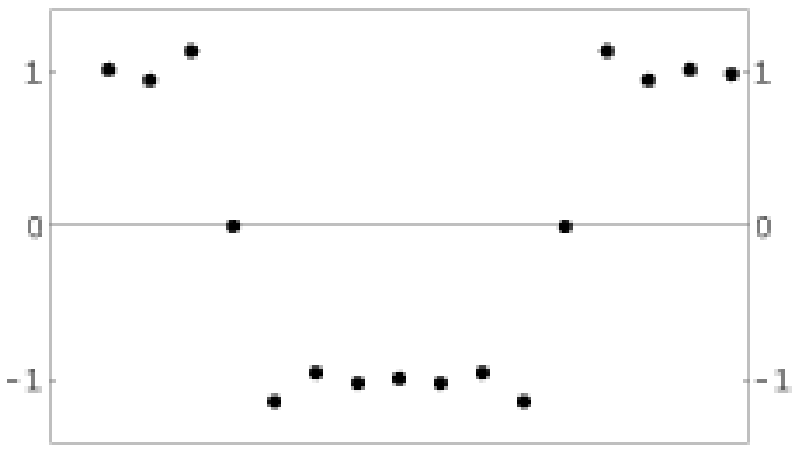}
\caption{%
Numerical solutions for the scalar field configuration 
in the cases with $C_4$ for the left column,
$C_8$ for the center column, and
$C_{16}$ for the right column. Each vertical axis
indicates
$y_i=\phi_i/a$. The figures in the five rows correspond to the parameter
$h=2^{-1}, 2^{-0.5}, 1, 2^{0.5}, 2^1$, from the top to the bottom.}
\label{fig3}
\end{figure}

FIG.~\ref{fig3} shows the solitonic profile
for $N=4, 8,~ {\rm and}~ 16$.
The feature is the same as the case with the path graph.

In both cases, a large $h$ means that the nonlinear part of the potential
dominates the difference-square part. 
The tendency indicates that the simultaneous limit $N\rightarrow \infty$
and
$h\rightarrow 0$ leads to the continuous solitons.

\section{Yukawa couplings
\label{fermion}}

Now we consider fermion fields coupled to the scalar fields.
Our starting point is the discretization
of the kinetic term in the continuum action in Sec.~\ref{kink}.
It is already known \cite{KS} that the fermion kinetic term can be
expressed in the incidence matrix ($E$). One set of chiral fermion fields
$\psi$ are assigned on vertices of a graph and the other set of chiral
fermion fields $\chi$ with the opposite chirality are assigned on edges
of a graph. Then the discretized kinetic term has the form
$$
-f\bar{\psi}E\chi+h.c.\,,
$$
where indices are suppressed.

Now we postulate three possibilities of Yukawa interaction.
We call them Case A, Case B, and Case C, for later convenience.

\subsection{a definition from difference operations
\label{difference}}

In the continuum theory, we have already seen that the scalar $(mass)^2$
matrix can be written in the form
\begin{equation}
\left|\frac{\partial W}{\partial \phi}\right|^2\,,
\end{equation}
in the two-dimensional $\phi^4$ theory in Sec.~\ref{kink}, while
in our model,
\begin{equation}
\left|\frac{{\it\Delta} W}{{\it\Delta} \phi}\right|^2\,,
\end{equation}
gives the potential, where
\begin{equation}
W(\phi_i)=\frac{\phi_i^3}{3}-a^2\phi_i\,.
\end{equation}

In the supersymmetric model, the
fermion mass matrix is often given in the form
\begin{equation}
\frac{\partial^2 W}{\partial \phi^2}\,,
\end{equation}
in the continuum theory.
The naive discretization of this form will be formally expressed as
\begin{equation}
\frac{{\it\Delta}^2 W}{{\it\Delta} \phi^2}\,.
\end{equation}

In actual, since ${\it\Delta} W/{\it\Delta} \phi$ is
\begin{equation}
g\left(
\frac{\phi_{i+1}^2+\phi_{i+1}\phi_i+\phi_i^2}{3}-a^2\right)\,,
\end{equation}
the difference of two sequential terms becomes
\begin{eqnarray}
& &g
\frac{\phi_{i+1}^2+\phi_{i+1}\phi_i-\phi_{i}\phi_{i-1}-\phi_{i-1}^2}{3}
=
g\frac{\phi_{i+1}+\phi_i+\phi_{i-1}}{3}(\phi_{i+1}-\phi_{i-1})
\,,\nonumber \\
&=&
g\frac{\phi_{i+1}+\phi_i+\phi_{i-1}}{3}(\phi_{i+1}-\phi_i+\phi_i-\phi_{i-1})
=g\frac{\phi_{i+1}+\phi_i+\phi_{i-1}}{3}({\it\Delta}\phi_{i}+{\it\Delta}\phi_{i-1})
\,.
\label{2nd}
\end{eqnarray}
Graph theoretically speaking, the difference ${\it\Delta}\phi$ lives on
each edge of a graph. 
Therefore we force chiral fermion fields on edges to couple the
second difference in (\ref{2nd}).
Namely, for a cycle graph, 
\begin{equation}
{\tiny
Y_d(C)=\frac{1}{3}\left(
\begin{array}{cccccc}
\phi_N+\phi_1+\phi_2& 0&0 &\cdots&0 &\phi_N+\phi_1+\phi_2\\
\phi_1+\phi_2+\phi_3&\phi_1+\phi_2+\phi_3&0 &\cdots&0&0\\
0& \phi_2+\phi_3+\phi_4&\phi_2+\phi_3+\phi_4&\cdots& 0&0\\
\vdots &\vdots &\vdots &\ddots&\vdots& \vdots\\
 0& 0&
0&\cdots&\phi_{N-2}+\phi_{N-1}+\phi_N&\phi_{N-2}+\phi_{N-1}+\phi_N\\
\phi_{N-1}+\phi_N+\phi_1& 0& 0&\cdots& 0&\phi_{N-1}+\phi_N+\phi_1
\end{array}
\right)\,,
}
\end{equation}
couples to $\chi$ as $Y_d(C)\chi$, where
\begin{equation}
\chi=\left(
\begin{array}{c}
\chi_1\\
\chi_2\\
\vdots\\
\chi_N
\end{array}
\right)\,.
\end{equation}

For a path graph, the coupling at both ends should be fixed by hand.
We take
\begin{equation}
{\tiny
Y_d(P)=\frac{1}{3}\left(
\begin{array}{cccccc}
2\phi_1+\phi_2& 0&0 &\cdots&0 &0\\
\phi_1+\phi_2+\phi_3&\phi_1+\phi_2+\phi_3&0 &\cdots&0&0\\
0& \phi_2+\phi_3+\phi_4&\phi_2+\phi_3+\phi_4&\cdots& 0&0\\
\vdots &\vdots &\vdots &\ddots&\vdots& \vdots\\
 0& 0&
0&\cdots&\phi_{N-2}+\phi_{N-1}+\phi_N&\phi_{N-2}+\phi_{N-1}+\phi_N\\
 0& 0& 0&\cdots& 0&\phi_{N-1}+2\phi_N
\end{array}
\right)
}
\end{equation}
which couples to $\chi$ as $Y_d(P)\chi$, where
\begin{equation}
\chi=\left(
\begin{array}{c}
\chi_1\\
\chi_2\\
\vdots\\
\chi_{N-1}
\end{array}
\right)\,.
\end{equation}

We consider the four-dimensional fermion bilinear term
\begin{equation}
f\bar{\psi}(-E+h'a^{-1}Y_d)\chi+h.c.\,,
\end{equation}
where $\psi=(\psi_1, \psi_2, \cdots, \psi_N)$ is fermions on vertices
of the opposite chirality to $\chi$. The coupling $h'$ is $Ga/f$, where
$G$ is the general Yukawa coupling.

\subsection{a definition from SUSY inspired structures
\label{susy}}

Suppose a cycle graph.
The superpotential $W_i$ can be defined on each edge as
\begin{equation}
W_i([\phi],[\Phi])=[\Phi]_i\left[f([\phi]_{i+1}-[\phi]_i)+g\left(
\frac{[\phi]_{i+1}^2+[\phi]_{i+1}[\phi]_i+[\phi]_i^2}{3}-a^2\right)\right]
\,,
\label{sac}
\end{equation}
where the superfields represent multiplets, each of which consists of 
a scalar, a chiral fermion, and an auxiliary field:
\begin{equation}
[\phi]_i\ni (\phi_i,\psi_i,F^v_i)\,,
\end{equation}
\begin{equation}
[\Phi]_i\ni(\Phi_i,\chi_i ,F^e_i)\,.
\end{equation}
Then we get
\begin{eqnarray}
\left(\frac{\partial W_i}{\partial \Phi_i}\right)^2&=&
f^2(\phi_{i+1}-\phi_i)^2+g^2\left(
\frac{\phi_{i+1}^2+\phi_{i+1}\phi_i+\phi_i^2}{3}-a^2\right)^2\nonumber
\\ & &+2fg\left(
\frac{\phi_{i+1}^3-\phi_i^3}{3}-a^2(\phi_{i+1}-\phi_i)\right)\,.
\end{eqnarray}
Assuming $\Phi_i=0$, the potential for $\phi_i$ is given by
\begin{equation}
\sum_{i=1}^N\left(\frac{\partial W_i}{\partial \Phi_i}\right)^2=
\sum_{i=1}^N\left[f^2(\phi_{i+1}-\phi_i)^2+\left(
\frac{\phi_{i+1}^2+\phi_{i+1}\phi_i+\phi_i^2}{3}-a^2\right)^2\right]\,,
\end{equation}
since there is no terminal vertex in a cycle graph.
This is just the scalar sector in our model considered in the present
paper.

Thus we can derive the coupling terms from (\ref{sac}).
The fermion bilinear term is found to be
\begin{equation}
\sum_i\bar{\chi}_i\left[f(\psi_{i+1}-\psi_i)+g
\frac{2\phi_{i+1}\psi_{i+1}+\phi_i\psi_{i+1}+\phi_{i+1}\psi_i
+2\phi_i\psi_i}{3}\right]+h.c.\,.
\end{equation}
Now we define
\begin{equation}
{\small
Y^T_s(C)=\frac{1}{3}\left(
\begin{array}{cccccc}
2\phi_1+\phi_2& \phi_1+2\phi_2&0 &\cdots&0 &0\\
0&2\phi_2+\phi_3&\phi_2+2\phi_3 &\cdots&0&0\\
\vdots &\vdots &\vdots &\ddots&\vdots& \vdots\\
 0& 0&
0&\cdots&2\phi_{N-1}+\phi_N&\phi_{N-1}+2\phi_N\\
\phi_N+2\phi_1& 0& 0&\cdots& 0&2\phi_N+\phi_1
\end{array}
\right)\,,
}
\end{equation}
and rewrite the bilinear term as
\begin{equation}
f\bar{\chi}(-E^T+ha^{-1}Y_s^T)\psi+h.c.\,.
\end{equation}
The coupling $h$ may be changed to $h'$ in a general case.

In the case with a path graph,
\begin{equation}
{\small
Y^T_s(P)=\frac{1}{3}\left(
\begin{array}{cccccc}
2\phi_1+\phi_2& \phi_1+2\phi_2&0 &\cdots&0 &0\\
0&2\phi_2+\phi_3&\phi_2+2\phi_3 &\cdots&0&0\\
\vdots &\vdots &\vdots &\ddots&\vdots& \vdots\\
 0& 0&
0&\cdots&2\phi_{N-1}+\phi_N&\phi_{N-1}+2\phi_N
\end{array}
\right)
}
\end{equation}
can be selected.
Note in this case, although the terminal vertices may
break an exact supersymmetry. Of course, supersymmetry is merely
a guiding principle here.

\subsection{a definition from a simple democracy on edges and vertices
\label{ed}}
Another simple choice respecting a graph structure for the Yukawa
couplings is possible. The definition is
\begin{equation}
{\small
Y^T_e(C)=\left(
\begin{array}{cccccc}
\phi_1& \phi_2&0 &\cdots&0 &0\\
0&\phi_2&\phi_3 &\cdots&0&0\\
\vdots &\vdots &\vdots &\ddots&\vdots& \vdots\\
 0& 0&
0&\cdots&\phi_{N-1}&\phi_N\\
\phi_1& 0& 0&\cdots& 0&\phi_N
\end{array}
\right)\,,
}
\end{equation}
for a cycle graph
and also
\begin{equation}
{\small
Y^T_e(P)=\left(
\begin{array}{cccccc}
\phi_1& \phi_2&0 &\cdots&0 &0\\
0&\phi_2&\phi_3 &\cdots&0&0\\
\vdots &\vdots &\vdots &\ddots&\vdots& \vdots\\
 0& 0&
0&\cdots&\phi_{N-1}&\phi_N
\end{array}
\right)
}
\end{equation}
for a path graph.

The four dimensional fermion bilinear term is
\begin{equation}
f\bar{\chi}(-E^T+h'a^{-1}Y_e^T)\psi\,.
\end{equation}

We considered three cases for Yukawa couplings.
In the next section, we will obtain the spectrum of the fermions for each
case.

\section{mass spectra for fermions and localization
\label{fmass}}

\begin{figure}[ht]
\centering
\includegraphics[height=3cm]
{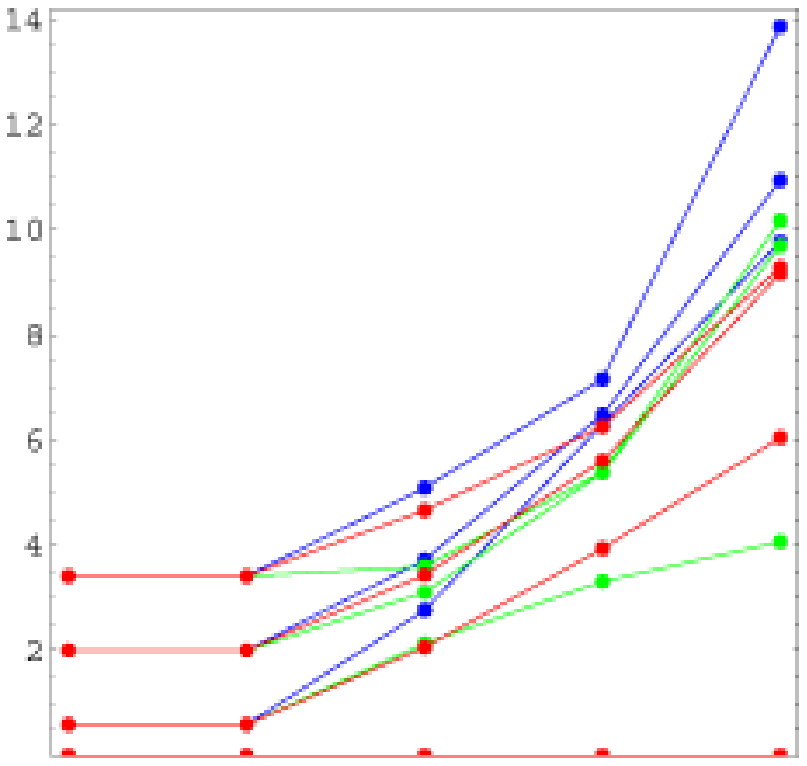}\qquad
\includegraphics[height=3cm]
{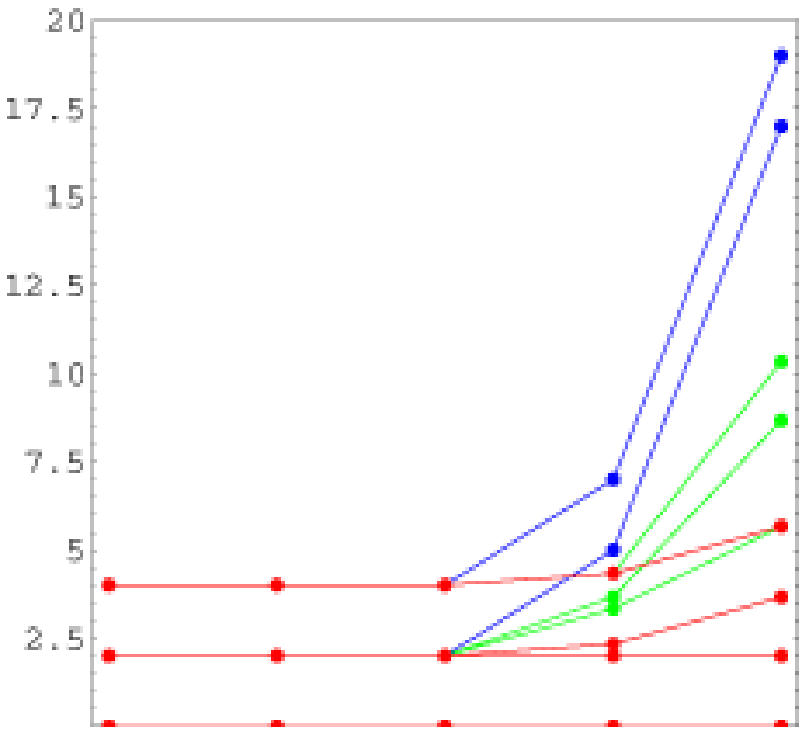}\\
\includegraphics[height=3cm]
{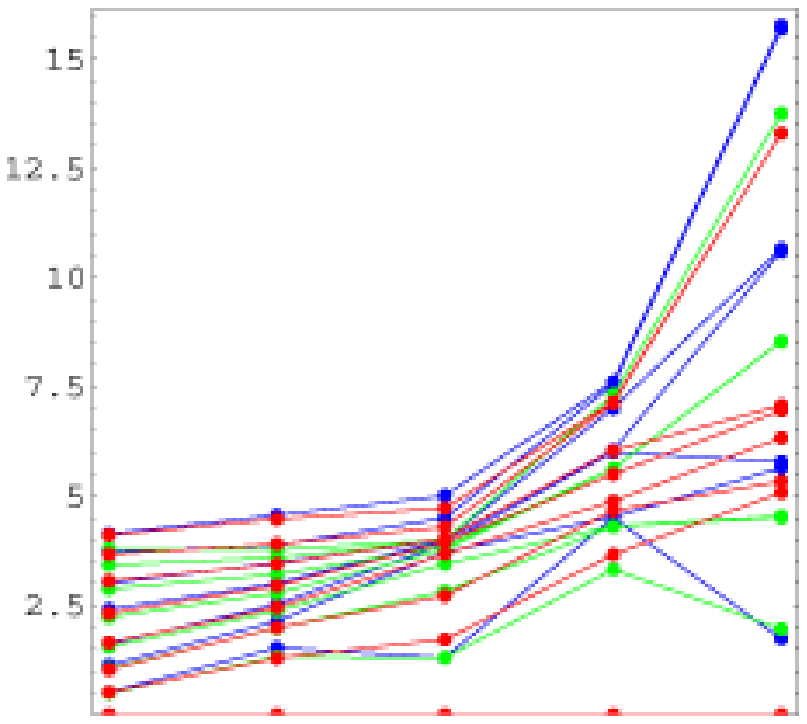}\qquad
\includegraphics[height=3cm]
{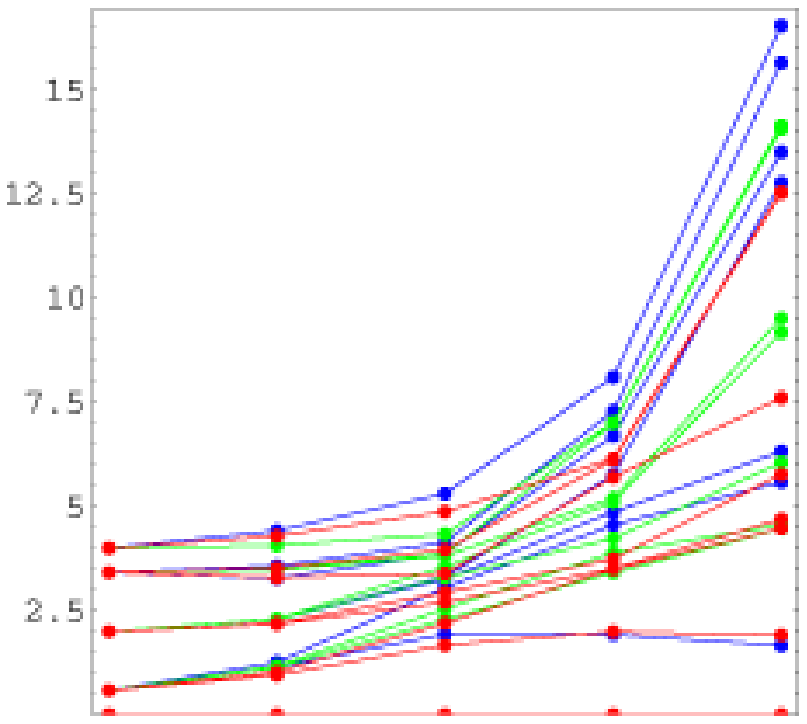}\\
\includegraphics[height=3cm]
{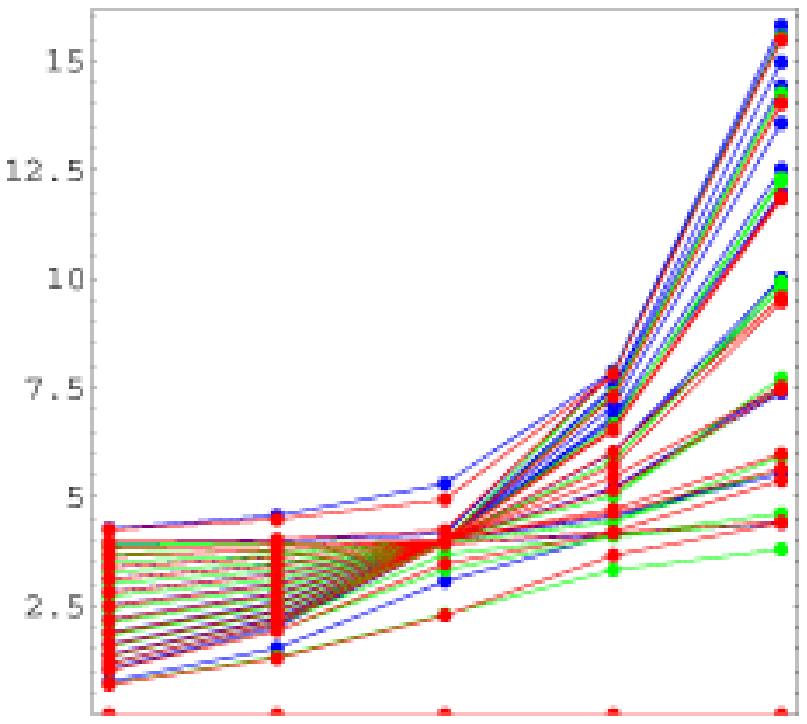}\qquad
\includegraphics[height=3cm]
{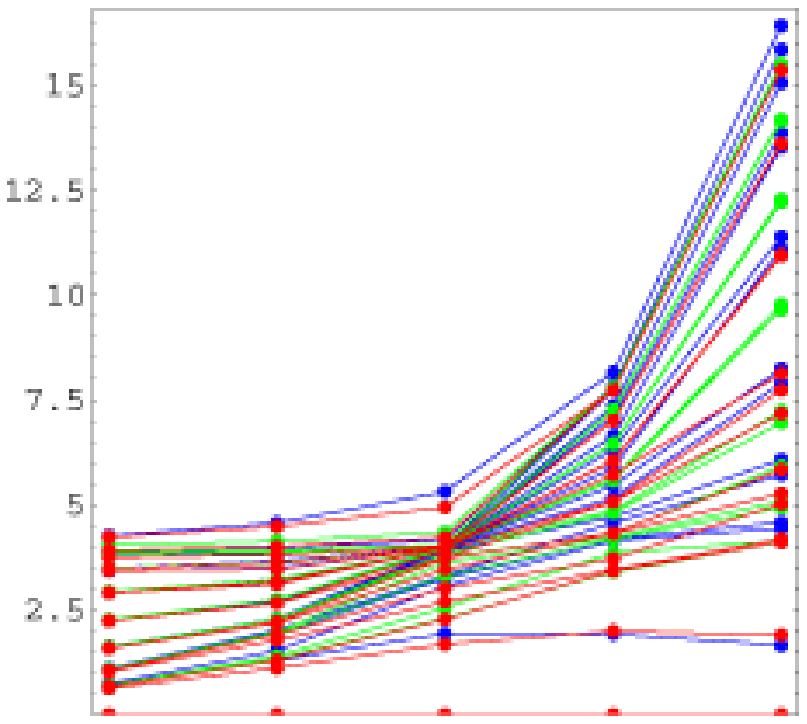}
\caption{%
The spectra of the fermion mass-square 
in the case with a path graph for the left column and
a cycle graph for the rught column. The figures in the three rows
correspond to 
$N=4, 8, 16$, from the top to the bottom.
The horizontal ticks correspond to the parameter
$h=2^{-1}, 2^{-0.5}, 1, 2^{0.5}, 2^1$, from the left to the right.
The red points indicate the eigenvalues in Case A, 
the green points indicate the eigenvalues in Case B, 
while the blue points indicate the eigenvalues in Case C. }
\label{fig4}
\end{figure}

FIG.~\ref{fig4} shows the fermion mass spectrum for the model associated
with a path graph and that with a cycle graph.
The eigenvalues of the matrix $(-E+ha^{-1}Y)(-E^T+ha^{-1}Y^T)$
are plotted, where the suffices indicating definitions of the Yukawa
coupling are suppressed. The Yukawa coupling constant is taken as $h'=h$,
so that the model in the continuum limit is expected to correspond with
the supersymmetric model in Sec.~\ref{kink}.

For the mode with a path graph $P_{16}$,
we can see the degeneracy of eigenvalues at $h\sim 1$.
In other cases, the degeneracy may occur at slightly large value for $h$.

It is known that the fermion spectrum has the finite discrete modes and
continuous modes in the background of a kink in the continuum
theory~\cite{DJR}.
The degeneracy of our case reflects the continuum spectrum in the
corresponding field theory.
This phenomenon bring about for the appropriate coupling $h$,
because the scalar configuration leaves far from the kink shape for small
and large $h$. For a small $h$, the difference-squared part is dominant.
For a large $h$, the non-linear potential is dominant.
In both cases the shape of a kink is deformed.
 
\begin{figure}[ht]
\centering
\includegraphics[height=3cm]
{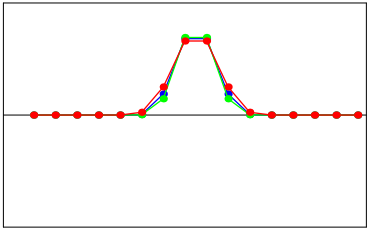}
\caption{%
The profiles of the eigenvectors associated with zero modes
for $P_{16}$ with $h=1$.
The red points indicate the eigenmodes in Case A, 
the green points indicate the eigenmodes in Case B, 
while the blue points indicate the eigenmodes in Case C. }
\label{loc}
\end{figure}

Finally we show the localization of zero mode fields.
In applications to the particle physics, the localization of matter
fields is expected and is substantial for inclusion of other interactions
among the matter fields~\cite{dw,ex}. In FIG.~\ref{loc}, the eigenvectors
associated with zero eigenvalues for
$P_{16}$ with $h=1$ are exhibited.

The difference of profiles of zero modes is minute for the three choices
for the Yukawa couplings.
As in the model with a continuous extra dimension,
we can utilize the `localization' for constructing elaborated gauge
interacting models with several gauge groups.

\section{Summary and outlook
\label{summary}}

We investigated the solitonic configuration of scalar fields 
in deconstructed theory, or in a discrete
extra space. It is expected that a large number of fields realizes the
thee configuration corresponding to the kink in the continuum space.

We found discrete eigenvalues for fermions appropriately coupled to the
kink-like configuration of scalar fields and the appearance of degeneracy
of them for a set of certain critical couplings.
This degeneracy is not observed in the continuum theory, where the
continuum spectrum can be found.
We also emphasis on the simple features of the
models:  the localization of
matter fields occurs and the existence of zero modes,
which correspond to the standard-model particles, is guaranteed.

The application of the deconstructed scalar model
to particle physics turns to be interesting. The degeneracy in the
spectrum may lead to the significant quantum effects on the zero mode
fields if the other interactions are turned on. The difference from the
higher-dimensional model with compactification is expected to be found,
since the continuum theory includes infinite tower of fields. Moreover
the study of the Casimir effect will be important especially if
gravitation is incorporated into the models. We think that the dynamical
creation and  evolution of the kink or `topological'  (not in the strict
meanings) configuration are worth studying in the cosmological context.

The generalization of models to those with the structure of general
graphs other than cycle and path graphs is  mathematically
interesting.  
We wish  also to investigate deconstructing non-linear sigma models
and `topological' configurations in them in the future works.

\begin{acknowledgments}
The authors would like to thank T.~Maki for critical reading this
manuscript.
\end{acknowledgments}



\bibliographystyle{apsrev4-1}

\end{document}